\documentclass[11pt,a4paper]{article}
\pdfoutput=1
\usepackage{jheppub}
\usepackage{epsfig,simplewick,xcolor,amsmath}
\usepackage{fancybox}

\def \Tr{\mbox{Tr\,}}

\newcommand{\be}{\begin{equation}}
\newcommand{\bea}{\begin{eqnarray}}
\newcommand{\ee}{\end{equation}}
\newcommand{\eea}{\end{eqnarray}}

\begin{document}
\title{Bosonic Fortuity in Vector Models}
\author[a,b]{Robert de Mello Koch\footnote{\href{mailto:robert@zjhu.edu.cn}{robert@zjhu.edu.cn}},}
\affiliation[a]{School of Science, Huzhou University, Huzhou 313000, China}
\affiliation[b]{Mandelstam Institute for Theoretical Physics, School of Physics, University of the Witwatersrand, Private Bag 3, Wits 2050, South Africa}
\author[a]{Animik Ghosh\footnote{Corresponding author; \href{mailto: animikghosh@gmail.com}{animikghosh@gmail.com}}}
\author[c]{Hendrik J.R. Van Zyl\footnote{\href{hjrvanzyl@gmail.com}{hjrvanzyl@gmail.com}}}
\affiliation[c]{The Laboratory for Quantum Gravity \& Strings,
Department of Mathematics \& Applied Mathematics,
University of Cape Town, Cape Town, South Africa}

\date{April 2025}
\abstract{We investigate the space of $U(N)$ gauge-invariant operators in coupled matrix-vector systems at finite $N$, extending previous work on single matrix models. By using the Molien-Weyl formula, we compute the partition function and identify the structure of primary and secondary invariants. In specific examples we verify, using the trace relations, that these invariants do indeed generate the complete space of gauge invariant operators. For vector models with $f \leq N$ species of vectors, the space is freely generated by primary invariants, while for $f > N$, secondary invariants appear, reflecting the presence of nontrivial trace relations. We derive analytic expressions for the number of secondary invariants and explore their growth. These results suggest a bosonic analogue of the fortuity mechanism. Our findings have implications for higher-spin holography and gauge-gravity duality, with applications to both vector and matrix models.}
\maketitle

\section{Introduction}

The structure of the space of gauge invariant operators for the multi-matrix quantum mechanics of $d$ matrices, at finite $N$, was recently considered in \cite{deMelloKoch:2025ngs}. For a single matrix, this space is the ring generated by single traces containing no more than $N$ matrices per trace. One might have expected a similar structure for multi-matrix models, but this naive expectation is incorrect. At finite $N$ there is a complete set of generating invariants, which fall into two distinct categories: \emph{primary} and \emph{secondary} invariants. Primary invariants are algebraically independent and generate the ring freely, while secondary invariants satisfy quadratic relations. The number of primary invariants grows as $\sim N^2$ which is much smaller that the exponentially larger growth of the secondary invariants $\sim e^{cN^2}$ with $c$ an order $1$ constant. Although all single-trace operators with $\leq N$ matrices appear among the generators, the generating set also includes invariants involving traces of more than $N$ matrices. There is no sense in which traces of matrices with $\leq N$ matrices represents a complete set.

In constructing the space of gauge-invariant operators, the primary invariants act freely and can appear with arbitrary multiplicity, naturally giving rise to a Fock space structure. They are therefore best interpreted as perturbative degrees of freedom. In contrast, each secondary invariant can appear only once and enters linearly, resembling a fixed background. From this perspective, secondary invariants serve as non-perturbative backgrounds that can be populated by perturbative excitations.\footnote{As demonstrated explicitly in~\cite{deMelloKoch:2025ngs}, most secondary invariants involve $O(N^2)$ fields. In the $\frac{1}{2}$-BPS sector of $\mathcal{N}=4$ super Yang–Mills theory, it is well known that under the AdS/CFT duality such operators correspond to new spacetime geometries~\cite{Lin:2004nb}. There are however some secondary invariants which involve only $O(N)$ fields. These are best thought of as new states that need to be added to obtain a complete description of the Hilbert space. There is again a complete Fock space built on each of them.} These gauge invariant operators are the collective fields in a collective field theory description \cite{Jevicki:1979mb,deMelloKoch:2002nq}.

The results in~\cite{deMelloKoch:2025ngs} were obtained by computing exact finite-$N$ partition functions using the Molien–Weyl formula. These partition functions take the form ($x=e^{-\beta}$)
\begin{equation}
Z(x) = \frac{1 + \sum_i c^s_i x^i}{\prod_j (1 - x^j)^{c^m_j}}, \label{illpf}
\end{equation}
which is precisely the structure predicted by the Hironaka decomposition~\cite{sturmfels,Grinstein:2023njq}. In this decomposition, the denominator captures the contribution of \emph{primary} invariants, while the numerator accounts for \emph{secondary} invariants. Specifically, $c^m_j$ counts the number of primary invariants constructed from $j$ fields, and $c^s_i$ counts the number of secondary invariants built from $i$ fields. As $N$ is increased, secondary invariants change their character and transition to become primary invariants. In the $N=\infty$ limit all invariants are primary. This matches the fact that at infinite $N$ there are no trace relations and the ring of gauge invariant operators is generated freely.

Finite-$N$ effects have also played a central role in recent developments in black hole microstate construction, particularly through the discovery of the \emph{fortuity mechanism}~\cite{Chang:2022mjp,Choi:2022caq,Choi:2023znd,Chang:2023zqk,Choi:2023vdm,Chang:2024zqi}. These $\frac{1}{16}$-BPS microstates fall into two classes: monotone and fortuitous. Monotone states preserve their BPS nature for all $N$, whereas fortuitous states cease to be BPS above a certain critical value of $N$. As a result, some states that appear as $\frac{1}{16}$-BPS microstates at a fixed $N$ no longer qualify as microstates when $N$ is increased. This breakdown is directly linked to trace relations, which play a crucial role in BPS saturation. Related discussions in the context of the SYK model and the D1–D5 system can be found in~\cite{Chang:2024lxt,Chang:2025rqy}.

A striking observation in~\cite{deMelloKoch:2025ngs} is that the growth in the number of secondary invariants at large $N$ closely tracks the expected scaling of black hole entropy. This suggests a natural interpretation: secondary invariants may correspond to black hole microstates. Moreover, the promotion of secondary to primary invariants as $N$ increases can be viewed as a purely bosonic analog of the fortuity mechanism.

An interesting direction in which the results of~\cite{deMelloKoch:2025ngs} can be extended is the study of coupled matrix--vector models. As we explain in Section~\ref{partfunc}, such models admit new types of gauge-invariant operator structures not present in purely matrix systems. In the context of gauge--gravity duality, this distinction reflects the fact that matrix models are expected to be dual to closed string theories, while coupled matrix--vector systems correspond to theories involving both open and closed strings.

In this letter, we explore the structure of the space of gauge-invariant operators at finite \( N \) in coupled matrix–vector systems. In Section~\ref{partfunc}, we derive the Molien–Weyl formula for the associated partition function. Section~\ref{simpleexample} examines a simple model with a single matrix and a single vector at \( N=2 \), where we show how trace relations allow the full space of gauge-invariant operators to be reconstructed from the generating set. In Section~\ref{VMF}, we turn to purely vector models. We demonstrate that the number of primary invariants is \( f^2 \) when \( f \le N \), and \( N^2 + 2N(f - N) \) when \( f > N \). In the latter case, secondary invariants appear, and we present evidence suggesting that their number grows asymptotically as \( e^{2\log 2\, Nf} \) for large \( f \). Section~\ref{HTL} analyzes the high-temperature limit of vector models, yielding an entropy \( S = f^2 \log T \) for \( f \le N \), and \( S = \text{const} + (N^2 + 2N(f - N)) \log T \) when \( f > N \). Section~\ref{M/V} extends the analysis to matrix–vector models. Here, we show how trace relations imply that secondary invariants are quadratically reducible, and we begin to quantify the number of primary and secondary invariants required to generate the complete ring of gauge-invariant operators. Finally, Section~\ref{Discussion} presents a discussion of our results and outlines potential directions for future research.

\section{Partition Function}\label{partfunc}

Consider the free quantum mechanics of $f$ complex vector fields $\phi^i_\alpha$, $\phi^{i\dagger}_\alpha$ for $i=1, 2,\ldots,f$, and $d$ Hermitian matrix fields $\Phi^I_{\alpha\beta}$ for $I=1, 2,\ldots,d$, described by the action
\begin{equation}
S = \int dt \sum_{i=1}^f \left( \partial_t \phi^{i\dagger} \cdot \partial_t \phi^i - \phi^{i\dagger} \cdot \phi^i \right) 
+ \int dt \sum_{I=1}^d \Tr\left( \partial_t \Phi^I \partial_t \Phi^I - \Phi^I \Phi^I \right),\label{oscillators}
\end{equation}
where the ``color'' indices $\alpha,\beta$ run over $1,2,\ldots,N$. The model has a global $U(N)$ symmetry, under which the fields transform as
\begin{equation}
\phi^i_{\alpha} \to U_{\alpha\beta} \phi^i_\beta, \qquad \phi^{i\dagger}_\alpha \to \phi^{i\dagger}_\beta U^\dagger_{\beta\alpha}, \qquad \Phi^I_{\alpha\beta} \to U_{\alpha\gamma} \Phi^I_{\gamma\tau} U^\dagger_{\tau\beta},
\end{equation}
i.e. $\phi^i$ is in the fundamental representation, $\phi^{i\dagger}$ is in the anti-fundamental and $\Phi^I$ is in the adjoint representation of $U(N)$.

We now promote this global $U(N)$ symmetry to a gauge symmetry, such that physical states are required to be gauge-invariant. This reduction to the singlet sector can be implemented by coupling the theory to a gauge field $A_0$, replacing ordinary derivatives with covariant derivatives. In one dimension $A_0$ is not dynamical and it simply imposes the Gauss Law as a constraint. The net effect is that the physical Hilbert space consists only of singlet states.

Gauge-invariant states are constructed from products of fields with all color indices fully contracted. There are two basic types of gauge-invariant operators: single-trace operators of the form $\Tr(W)$, and bilinears of the form $\phi^{i\dagger}\cdot W\cdot \phi^j$, where $W$ is any word built using the matrices $\Phi^I$ as letters. More general gauge invariant operators are constructed as arbitrary polynomials of these two basic types. Our goal is to determine the complete ring of such gauge invariant operators at finite $N$ and to understand its algebraic structure. 

Towards this end, we would like to compute the partition function of these theories. To keep track of the various fields of the theory, we introduce three sets of chemical potentials $\mu_{\phi^i}$, $\mu_{\phi^{i\dagger}}$ and $\mu_{\Phi^I}$ and we note that the oscillators defined by (\ref{oscillators}) all have unit energy spacing. The partition function is written in terms of the variables
\bea
x^i=e^{-\beta E_{\phi^i} -\mu_{\phi^i}}\qquad
y^i=e^{-\beta E_{\phi^{i\dagger}} -\mu_{\phi^{i\dagger}}}\qquad
z^I=e^{-\beta E_{\Phi^I}-\mu_{\Phi^I}}
\eea
Since we have a free theory the spectrum is additive and the exact partition function is\footnote{Very helpful references are \cite{Sundborg:1999ue,Aharony:2003sx} as well as \cite{Amado:2017kgr}.}
\bea
Z(x^i,y^i,z^I)&=&\prod_{i=1}^j\prod_{I=1}^d\sum_{n_i=0}^\infty\sum_{m_i=0}^\infty\sum_{p_I=0}^\infty (x^i)^{n_i}(y^i)^{m_i}(z^I)^{p_i}\times \#(\{n_i,m_i,p_I\})
\eea
where $\#(\{n_i,m_i,p_I\})$ is the number of singlets that can be constructed using $n_i$ $\phi^i$ fields, $m_i$ $\phi^{i\dagger}$ fields and $p_I$ $\Phi^I$ fields. Using characters we can write the number of singlets as an integral over $U(N)$ to obtain
\bea
Z(x^i,y^i,z^I)&=&\int_{U(N)} [DU]\prod_{i=1}^j\sum_{n_i=0}^\infty (x^i)^{n_i}\chi_{{\rm sym}^{n_i}(F)}(U)\sum_{m_i=0}^\infty (y^i)^{m_i}\chi_{{\rm sym}^{m_i}(\bar{F})}(U)\cr\cr
&&\qquad\times\prod_{I=1}^d \sum_{p_I=0}^\infty (z^I)^{p_I}\chi_{{\rm sym}^{p_I}(adj)}(U)
\eea
$F,\bar{F}$ stand for the fundamental and anti-fundamental representations, while $adj$ stands for the adjoint representation. The reason why the symmetric product of these representations appears is that the fields we are using are all bosonic, so they must be symmetric under any permutation of fields of the same species. After making use of the identity
\bea
\sum_{n=0}^\infty x^n \chi_{{\rm sym}^{n}(R)}(U)={\rm exp}\left(\sum_{m=1}^\infty {x^m\over m}\chi_R(U^m)\right)
\eea
the partition function becomes
\bea
Z(x^i,y^i,z^I)&=&\int_{U(N)} [DU]\,\,
{\rm exp}\left(\sum_{i=1}^f \sum_{m=1}^\infty {(x^i)^m\over m}\chi_F(U^m)+\sum_{i=1}^f \sum_{m=1}^\infty {(y^i)^m\over m}\chi_{\bar{F}}(U^m)\right)\cr\cr
&&\qquad {\rm exp}\left(\sum_{I=1}^d \sum_{m=1}^\infty {(z^I)^m\over m}\chi_{adj}(U^m)\right)
\eea
Notice the the integrand above depends only on the eigenvalues $\varepsilon_i$ of $U$. Recall that $|\varepsilon_i|=1$ and $\varepsilon_i^*=\varepsilon_i^{-1}$. Using the explicit expressions for the characters in terms of the eigenvalues
\bea
\chi_F(U^m)&=&\sum_{i=1}^N (\varepsilon_i)^{m}\qquad\qquad \chi_{\bar{F}}(U^m)\,\,=\,\,\sum_{i=1}^N (\varepsilon_i)^{-m}\qquad
\chi_{\rm adj}(U^{\dagger\,m})\,\,=\,\,\sum_{i,j=1}^N (\varepsilon_i)^{m}(\varepsilon_j)^{-m}\nonumber
\eea
and the fact that for an integrand that depends only on the eigenvalues we can replace\footnote{Separating the parameters of the unitary matrix $U$ into eigenvalues and angles, we derive this replacement by integrating over the angles, using the fact that the integrand depends only on the eigenvalues.} 
\bea
\int_{U(N)} [DU] &\to& {1\over N! (2\pi i)^N}\oint \prod_{j=1}^N {d\varepsilon_j\over\varepsilon_j}\Delta\bar{\Delta}
\eea
where the Van der Monde determinants are given by
\bea
\Delta&=&\prod_{k<r}(\varepsilon_r-\varepsilon_k)\,\,=\,\,\sum_{\sigma\in S_N}{\rm sgn}(\sigma)\varepsilon^0_{\sigma(1)}\varepsilon^1_{\sigma(2)}\cdots\varepsilon^{N-1}_{\sigma(N)}\label{VdM}
\eea
\bea
\bar{\Delta}&=&\prod_{k<r}(\varepsilon_r^{-1}-\varepsilon_k^{-1})\,\,=\,\,\sum_{\tau\in S_N}{\rm sgn}(\tau)\varepsilon^0_{\tau(1)}\varepsilon^{-1}_{\tau(2)}\cdots\varepsilon^{-N+1}_{\tau(N)}\label{barVdM}
\eea
Using the above results, the partition function becomes
\bea
Z(x^i,y^i,z^I)&=&{1\over N! (2\pi i)^N}\oint \prod_{j=1}^N {d\varepsilon_j\over\varepsilon_j}\,\,\Delta\bar{\Delta}\,\,\prod_{l=1}^f\prod_{k=1}^N {1\over (1-x^l\varepsilon_k)(1-y^l \varepsilon^{-1}_k)}\cr\cr
&&\times \prod_{I=1}^d{1\over (1-z^I)^N}\prod_{1\le k<r\le N}{1\over (1-{\varepsilon_r\over\varepsilon_k}z^I)(1-{\varepsilon_k\over\varepsilon_r}z^I)}
\eea
We now simplify the integrand of the above formula. According to (\ref{VdM}) and (\ref{barVdM}) the product $\Delta\bar{\Delta}$ involves a sum over permutations $\sigma$ and $\tau$. For each distinct term in the sum over $\sigma$, consider the distinct change of variables given by a permutation of the eigenvalues $\varepsilon_{i}\to \varepsilon_{\sigma(i)}$. All factors in the integrand are invariant under this change except for the Van der Monde determinants which are both either invariant or both pick up a minus sign. Consequently the product $\Delta\bar{\Delta}$ is invariant. The net result of this change of variables is that the sum over $\sigma$ can be performed and we find
\bea
\Delta\bar{\Delta}\to  N!\varepsilon^0_1\varepsilon^1_2\cdots\varepsilon^{N-1}_{N}
\prod_{k<r}(\varepsilon_r^{-1}-\varepsilon_k^{-1})
\eea
It is now useful to change variables to the new variables $t_j$ defined by $\varepsilon_j=t_1t_2\cdots t_j$. A straightforward computation proves that
\bea
N!\varepsilon^0_1\varepsilon^1_2\cdots\varepsilon^{N-1}_{N}
\prod_{k<r}(\varepsilon_r^{-1}-\varepsilon_k^{-1})&=&N!\prod_{2\le k\le r\le N}(1-t_{k,r})
\eea
and
\bea
\prod_{1\le k<r\le N}\left(1-{\varepsilon_r\over\varepsilon_k}z\right)\left(1-{\varepsilon_k\over\varepsilon_r}z\right)&=&\prod_{2\le k\le r\le N}(1-zt_{k,r})(1-zt_{k,r}^{-1})
\eea
where $t_{k,r}=t_kt_{k+1}\cdots t_{r-1}t_r$. The Jacobian of this transformation is
\bea
J=\det {\partial\varepsilon_i\over \partial t_j}=\prod_{j=1}^N {\varepsilon_j\over t_j}
\eea
Thus, the partition function finally becomes
\bea
Z(x^i,y^i,z^I)&=&{1\over (2\pi i)^N}\oint \prod_{j=1}^N {dt_j\over t_j}\, \, \prod_{i=1}^f\,\prod_{l=1}^N {1\over (1-x^i t_{1,l})(1-{y^i \over t_{1,l}})}\cr\cr
&&\qquad\times \prod_{2\le k\le r\le N}\frac{1-t_{k,r}}{f_{k,r}}\prod_{I=1}^d {1\over (1-z^I)^{N}}\label{finalPF}
\eea
where
\bea
f_{k,r}\,\,=\,\,\prod_{I=1}^d\,(1-z^I t_{k,r})(1-z^I t_{k,r}^{-1})
\eea
After fixing a specific value of $N$ in (\ref{finalPF}), the resulting expression is used to evaluate the exact partition function. Each integral over the variables \( t_i \) is performed along the unit circle in the complex plane and can be evaluated using the residue theorem. As a consistency check, setting $x^i=y^i=0$ reproduces the results obtained in~\cite{deMelloKoch:2025ngs}, providing a non-trivial validation of our formula. Alternatively, by setting $z^I = 0$, we recover the partition function for a model that contains only vector fields and no matrix degrees of freedom.

\section{Warm up: one matrix, one vector and $N=2$}\label{simpleexample}

A simple model which nicely illustrates the physics we wish to explore is given by setting $d=1=f$ and $N=2$. The exact partition function is obtained by performing the double integral
\bea
Z(x,y,z)&=&{1\over (2\pi i)^2}\oint {dt_1\over t_1}\oint {dt_2\over t_2}\frac{1-t_2}{(1-t_1 x) \left(1-\frac{y}{t_1}\right) (1-t_1 t_2 x) \left(1-\frac{y}{t_1 t_2}\right)}\cr\cr
&&\qquad\times\frac{1}{(1-z)^2 \left(1-\frac{z}{t_2}\right) (1-t_2 z)}
\eea
For the integral over $t_1$ pick up the residues at $y$ and ${y\over t_2}$, and for the subsequent integral over $t_2$ pick up the residue at $xy$ and $z$. The result is
\bea
Z(x,y)={1\over (1-xy)(1-z)(1-z^2)(1-xyz)}
\eea
There are four primary invariant and no non-trivial secondary invariants. The primary invariants are $\phi^\dagger\cdot\phi$, $\Tr(\Phi)$, $\Tr(\Phi^2)$ and $\phi^\dagger\cdot\Phi\cdot\phi$. 

Following the arguments presented in~\cite{deMelloKoch:2025ngs}, the result above implies that the full space of gauge-invariant operators is freely generated by the four identified primary invariants. To rigorously establish this claim, one must use the trace relations that arise at finite $N$. For this purpose, it is useful to define the matrix $\Psi_{\alpha\beta}= \phi_\alpha \phi^\dagger_\beta$. The complete set of trace relations is then obtained from the Cayley–Hamilton theorem, which, for $N = 2$, yields the identity
\begin{align}
&\Tr(A)\Tr(B)\Tr(C) - \Tr(AB)\Tr(C) - \Tr(AC)\Tr(B) - \Tr(A)\Tr(BC) \notag \\
&\qquad + \Tr(ABC) + \Tr(ACB) = 0,
\end{align}
where $A$, $B$ and $C$ are arbitrary words formed from $\Phi$ and $\Psi$ as letters. In our analysis, it will suffice to consider trace relations involving only a single $\Psi$.

We begin by proving that $\Tr(\Phi^{n+1})$ can be constructed from the four invariants, assuming we can already construct all $\Tr(\Phi^k)$ for $k \leq n$. Since $\Tr(\Phi)$ and $\Tr(\Phi^2)$ are among the generators, this inductive step suffices to prove that $\Tr(\Phi^n)$ can be constructed for all $n \geq 3$. The required recursion follows by setting $A=\Phi^{n-1}$ and $B=C=\Phi$ in the trace relation, yielding
\begin{equation}
\Tr(\Phi^{n+1}) = \frac{1}{2} \left( 2 \Tr(\Phi^n) \Tr(\Phi) + \Tr(\Phi^{n-1}) \Tr(\Phi^2) - \Tr(\Phi^{n-1}) \Tr(\Phi)^2 \right).
\end{equation}

A completely analogous argument applies to the bilinear invariants $\phi^\dagger\cdot\Phi^n \cdot\phi$. The recursion relation needed follows by setting $A=\Phi^n$, $B=\Psi$ and $C=\Phi$ in the trace relation, which gives
\begin{align}
\phi^\dagger \cdot \Phi^{n+1} \cdot \phi &= \frac{1}{2} \Big( \phi^\dagger \cdot \Phi^n \cdot \phi \, \Tr(\Phi) + \phi^\dagger \cdot \phi \, \Tr(\Phi^{n+1}) \notag \\
&\qquad + \phi^\dagger \cdot \Phi \cdot \phi \, \Tr(\Phi^n) - \phi^\dagger \cdot \phi \, \Tr(\Phi^n) \Tr(\Phi) \Big).
\end{align}
This completes the proof that all gauge-invariant operators can be generated from the four primary invariants, confirming the result.

\section{Vector Model Fortuity}\label{VMF}

In this section, we focus on models containing only vector fields, i.e., we set $d=0$. The form of the partition function in these models exhibits a qualitative difference depending on whether $f\leq N$ or $f>N$. When $f\leq N$, the partition function takes the simple form
\begin{equation}
Z(x^i, y^i) = \prod_{i,j=1}^f \frac{1}{1 - x^i y^j},
\end{equation}
which corresponds to a freely generated ring of invariants. In this case, there are a total of $f^2$ primary invariants, with no non-trivial secondary invariants present.

Now consider the case where $f>N$. A simple illustrative example is given by $N=2$ and $f=3$. In such cases, the partition function graded by individual field species does not admit a Hironaka decomposition. This signals that the generating invariants for the space of gauge-invariant operators do not preserve the species grading \cite{Dokovic}. Consequently, we shift our focus to the \emph{blind} partition function, which does not distinguish between vector species. This is achieved by setting $x^1 = x^2 = x^3 = x$ and $y^1 = y^2 = y^3 = y$. The resulting partition function is
\begin{equation}
Z(x, y) = \frac{1 + x y + x^2 y^2}{(1 - x y)^8}.
\end{equation}
From this expression, we deduce that the space of gauge-invariant operators is generated by 8 primary invariants and 2 non-trivial secondary invariants. Notice further that the numerator of this partition function is palindromic.

There are several important points to clarify. First, the change in behavior of the partition function precisely coincides with the point at which trace relations begin to play a role. In a theory with a $U(N)$ gauge symmetry, non-trivial trace relations require at least $N+1$ independent indices that can be antisymmetrized. This condition translates into the requirement $f\ge N+1$ for such relations to appear. 

Explicit evaluation reveals that the partition function takes the general form
\begin{equation}
Z(x, y) = \frac{P(x, y)}{(1 - x y)^{N^2 + 2N(f - N)}},
\end{equation}
implying that the number of primary invariants is always given by $N^2+2N(f-N)$. This result can be understood through a simple counting argument. To make this concrete, consider the case $N=4$ and $f=6$. Using the $U(4)$ gauge symmetry, we can bring the vectors into the form
\begin{equation}
\phi^1 = \begin{bmatrix} r_1 \\ 0 \\ 0 \\ 0 \end{bmatrix}, \quad
\phi^2 = \begin{bmatrix} c_1 \\ r_2 \\ 0 \\ 0 \end{bmatrix}, \quad
\phi^3 = \begin{bmatrix} c_2 \\ c_3 \\ r_3 \\ 0 \end{bmatrix}, \quad
\phi^4 = \begin{bmatrix} c_4 \\ c_5 \\ c_6 \\ r_4 \end{bmatrix}, \quad
\phi^5 = \begin{bmatrix} c_7 \\ c_8 \\ c_9 \\ c_{10} \end{bmatrix}, \quad
\phi^6 = \begin{bmatrix} c_{11} \\ c_{12} \\ c_{13} \\ c_{14} \end{bmatrix}.\label{gaugefixed}
\end{equation}
We begin by using $U(4)$ to fix three components of $\phi^1$ to zero and make the remaining entry real and positive. Then, the $U(3)$ subgroup that preserves $\phi^1$ is used to fix two components of $\phi^2$ and render one component real and positive. Similarly, the $U(2)$ subgroup fixing both $\phi^1$ and $\phi^2$ is used to eliminate one component of $\phi^3$ and make another real and positive. Finally, the $U(1)$ subgroup preserving $\phi^1$, $\phi^2$, and $\phi^3$ is used to fix one component of $\phi^4$ to be real and positive. This process fully fixes the $U(4)$ gauge symmetry and leaves $32$ independent real parameters in the vector degrees of freedom. Carrying out this counting procedure for general $N$ and $f$ leads to the expression $N^2+2N(f-N)$ for the number of independent gauge-invariant degrees of freedom, which matches the exponent in the denominator of the partition function and hence the number of primary invariants. This argument counts the number of independent components of the vector fields that remain once the $U(N)$ symmetry is completely accounted for. This is not the same as counting the number of invariants. Since it might not be manifestly obvious, we explain why these two counting problems are identical in Appendix \ref{relatecount}.

\begin{table}[h]
\centering
\begin{tabular}{|c|c|c|c|}
\hline
 & \textbf{$N=2$} & \textbf{$N=3$} & \textbf{$N=4$} \\
\hline
\textbf{$f=2$} & 1 & 1 & 1 \\
\hline
\textbf{$f=3$} & 3 & 1 & 1 \\
\hline
\textbf{$f=4$} & 20 & 4 & 1 \\
\hline
\textbf{$f=5$} & 175 & 50 & 5 \\
\hline
\textbf{$f=6$} & 1764 & 980 & 105 \\
\hline
\textbf{$f=7$} & 19404 & 24696 & 4116 \\
\hline
\textbf{$f=8$} & 226512 & 731808 & 232848 \\
\hline
\textbf{$f=9$} & 2760615 & 24293412 & 16818516 \\
\hline
\end{tabular}
\caption{Count of the number of secondary invariants for specified $N$ and $f$ with the identity included in the count.}
\label{tab:secinv}
\end{table}

Although we have not been able to derive a formula for the number of secondary invariants, we have studied several examples (summarized in Table \ref{tab:secinv}) that allow us to identify the counting with some known integer sequences. The $N=2$ column matches the sequence A000891 in OEIS. From that entry, the analytic formula for the number of secondary invariants is 
\bea
N_{\rm secondary}&=&{(2f-2N)!(2f-2N+1)!\over \left((f-N)!(f-N+1)!\right)^2}\, .\label{Nmsc}
\eea
In the large $f$ limit this behaves as
\bea
N_{\rm secondary}&\approx&e^{4\log 2 \,\,f}
\eea
The $N=3$ column matches the sequence A282136 in OEIS. From that entry, the analytic formula for the number of secondary invariants is
\bea
N_{\rm secondary}&=&\frac{4 (2f-2N)! ((2 f-2N+1)!)^2}{((f-N)!)^3 (f-N+1)! ((f-N+2)!)^2}
\eea
In the large $f$ limit this behaves as
\bea
N_{\rm secondary}&\approx&e^{6\log 2 \,\,f}
\eea
The $N=4$ column does not match any of the sequences in OEIS. However, based on the series above we have managed to obtain the following analytic formula for the number of secondary invariants
\bea
N_{\rm secondary}&=&\frac{24 (2f-2N)! ((2f-2N+1)!)^2 (2f-2N+3)!}{((f-N)!)^3 (f-N+1)! ((f-N+2)!)^2 ((f-N+3)!)^2}
\eea
In the large $f$ limit this behaves as
\bea
N_{\rm secondary}&\approx&e^{8\log 2 \,\,f}
\eea
The formula that fits all of these examples is\footnote{Although we do not quote the numbers, we have also tested this formula for $N=5,6$.} 
\bea
N_{\rm secondary}&=&\prod_{j=1}^N\frac{(2(f-N)+j-1)!}{(((f-N)+j-1)!)^2}(j-1)!
\eea
The growth for large $f$ and fixed $N$ is $N_{\rm secondary}\approx e^{2\log 2 \,\,Nf}$. Thus, if we hold $N$ fixed and take $f\to\infty$ the number of secondary invariants grows as the exponential of $f$. Exactly the same growth was observed for the number of states in a collective field theory treatment of the singlet sector of the $Sp(2N)$ sigma model~\cite{Das:2012dt}. The collective fields of this description are bilocals, matching the form of the invariants counted by our partition function.

We also observe that all of the partition functions computed in this work are palindromic. As an illustrative example, consider the case $N=3$ and $f=7$, for which the partition function takes the form
\begin{equation}
Z(x, y) = \frac{P_{N=3, f=7}(x, y)}{(1 - x y)^{33}},
\end{equation}
where the numerator polynomial is given by
\begin{align}
P_{N=3, f=7}(x, y) &= 1 + 16 x y + 136 x^2 y^2 + 816 x^3 y^3 + 2651 x^4 y^4 + 5312 x^5 y^5 + 6832 x^6 y^6 \notag \\
&\quad + 5312 x^7 y^7 + 2651 x^8 y^8 + 816 x^9 y^9 + 136 x^{10} y^{10} + 16 x^{11} y^{11} + x^{12} y^{12}.
\end{align}
The palindromic nature of these partition functions follows directly from the structure of the Molien--Weyl formula~\cite{3mats}. Specifically, see Proposition 2.3 in~\cite{3mats}, and also~\cite{Kristensson:2020nly} for related discussion. The key idea underlying this property is intuitive: under the substitution $x \to 1/x$, $y \to 1/y$, the partition function transforms into itself up to an overall factor, as a consequence of the fact that $P_{N=3, f=7}(x, y)$ is palindromic. This transformation law can be seen directly from the Molien--Weyl integral by making the change of integration variables $t_i\to 1/t_i$. Since the integral is performed over unit circles in the complex plane, this substitution exchanges the poles inside and outside the contour. By Cauchy's residue theorem, the total sum of residues vanishes, implying that the contributions from the residues inside the contour equal (up to a sign) those from the outside. 

The palindromic property of the partition function reflects a deeper symmetry of the system: a $T \to -T$ invariance, as discussed in~\cite{McGady:2017rzv}. This symmetry provides a compelling structural constraint on the spectrum of gauge-invariant operators and is expected on general theoretical grounds. The fact that our partition functions are all palindromic is a nice consistency check of our results.

As we have seen above as soon as $f>N$, secondary invariants appear among the generators of the gauge-invariant operator ring. As $N$ increases, the number of secondary invariants gradually decreases, vanishing entirely once $N = f$. This behavior suggests that, even in purely bosonic vector models, there exists an analogue of the fortuity mechanism: secondary invariants, which are initially non-trivial, are promoted to primary status as the rank of the gauge group increases. 

\section{High Temperature Limit}\label{HTL}

For the vector model, we now have a good understanding of the structure of the partition function as a function of $N$ and $f$. Let us focus on the regime $f>N$, where the partition function takes the form
\begin{equation}
Z(x, y) = \frac{P(x, y)}{(1 - x y)^{N^2 + 2N(f - N)}},
\end{equation}
with $P(x, y)$ a polynomial containing only positive terms. To analyze the high-temperature behavior, we set the chemical potentials to zero so that $x=y=e^{-\beta}$. The high temperature limit $\beta\to 0$ corresponds to $x\to 1$. Since $P(x, y)$ is a sum of positive terms, it approaches a constant value in this limit: $P_{N, f} \equiv P(1, 1)$. To obtain this result, it is necessary that $T\to\infty$ rapidly enough that the highest-degree terms in $P(x, y)$ also approach 1. Since the highest degree of $P(x, y)$ will generally scales with $N$, this implies that $T$ must grow as a power of $N$. Determining this scaling precisely is an open problem, but it will ultimately dictate the precise high-temperature behavior.

Next, consider the leading behavior of the denominator in this limit:
\begin{equation}
(1 - x y)^{N^2 + 2N(f - N)} = \left(1 - e^{-2\beta}\right)^{N^2 + 2N(f - N)} \to 2^{N^2 + 2N(f - N)} T^{-N^2 - 2N(f - N)},
\end{equation}
where we have used $T=1/\beta$ and the approximation $1-e^{-2\beta}\approx 2\beta$ as $\beta\to 0$. Thus, in the high-temperature limit, the partition function simplifies to
\begin{equation}
Z(T) \approx \frac{P_{N, f}}{2^{N^2 + 2N(f - N)}} T^{-N^2 - 2N(f - N)}.
\end{equation}
The corresponding free energy is
\begin{equation}
F = -T \log Z(T) = -T \log \left( \frac{P_{N, f}}{2^{N^2 + 2N(f - N)}} \right) - \left(N^2 + 2N(f - N)\right) T \log T,
\end{equation}
and the entropy $S$ follows from the thermodynamic relation
\begin{equation}
S = -\frac{\partial F}{\partial T},
\end{equation}
yielding
\begin{equation}
S = \log \left( \frac{P_{N, f}}{2^{N^2 + 2N(f - N)}} \right) + \left(N^2 + 2N(f - N)\right) \log T.
\end{equation}
A similar analysis applies in the case $f\leq N$. In that regime, the number of primary invariants is simply $f^2$, and the high-temperature behavior of the entropy reduces to
\begin{equation}
S = f^2 \log T.
\end{equation}

\section{Matrix/Vector systems}\label{M/V}

For models with a single matrix, it is well-known that there are \( N \) primary invariants and no secondary invariants. The partition function in this case is given by
\begin{equation}
Z(z) = \prod_{a=1}^N \frac{1}{1 - z^a}.
\end{equation}

In the case of a model with a single vector ($f=1$) and a single matrix ($d=1$), the partition function takes the form
\begin{equation}
Z(x^i, y^i, z) = \prod_{a=1}^N \frac{1}{1 - z^a} \prod_{b=0}^{N-1} \frac{1}{1 - x z^b y},
\end{equation}
In this model, the number of primary invariants is $2N$, and there are no non-trivial secondary invariants.

For a model with $f = 2$ flavors, we encounter new structural features. Taking $N=2$, $d=1$, and $f=2$, the exact partition function is given by
\begin{equation}
Z(x, y, z) \,\,=\,\,\frac{1+xyz-x^2y^2z-x^3y^3z^2}{(1 - z)(1 - z^2)(1 - x y)^4(1 - x y z)^3},
\end{equation}
which is not of the Hironaka form due to the presence of both positive and negative terms in the numerator. We have not found a way to rewrite this partition function to bring it into Hironaka form while preserving the full vector/matrix grading. However, if we discard this grading by substituting $x y \rightarrow z$ -- which effectively treats a bilinear in the vector fields as having the same weight as a single matrix -- the partition function simplifies to
\begin{equation}
Z(z) \,\,=\,\, \frac{1 + z + 2 z^2 + z^3 + z^4}{(1 - z)^4 (1 - z^2)^4},
\end{equation}
which \emph{is} of the Hironaka form and it indicates that the space of gauge invariant operators is generated by 8 primary invariants and 5 non-trivial secondary invariants. The failure of the original graded partition function to admit a Hironaka decomposition suggests that at least one of the generators of the invariant ring must mix vector and matrix degrees of freedom. In other words, some invariants cannot be classified as purely matrix or purely vector but instead arise as nontrivial sums of both types.

When we include more than a single matrix, secondary invariants naturally emerge as they are already present in the pure multi-matrix problem. For instance, when $N=2$, including two species of matrices and a single vector leads to the following form for the partition function
\bea
Z(z^1, z^2, x, y) &=& \frac{1}{(1-z^1)(1-z^2)(1-(z^1)^2)(1-(z^2)^2)(1-z^1 z^2)}\cr\cr &&\times \frac{1 + x z^1 z^2 y}{(1 - xy)(1 - x z^1 y)(1 - x z^2 y)}.
\eea
In this case, there are 8 primary invariants. There are 5 single trace operators
\bea
m_1&=&\Tr(\Phi^1)\qquad m_2\,\,=\,\,\Tr(\Phi^2)\cr\cr
m_3&=&\Tr(\Phi^1 \Phi^1)\qquad m_4\,\,=\,\,\Tr(\Phi^2 \Phi^2)\qquad
m_5\,\,=\,\,\Tr(\Phi^1 \Phi^2)
\eea
as well as 3 mixed matrix/vector invariants
\bea
m_6&=&\phi^\dagger\cdot \phi\qquad m_7\,\,=\,\,\phi^\dagger\cdot \Phi^1\cdot\phi\qquad
m_8\,\,=\,\,\phi^\dagger\cdot\Phi^2\cdot\phi .
\eea
There is also a single non-trivial secondary invariant given by 
\bea
s&=&\phi^\dagger\cdot\Phi^1\Phi^2\cdot\phi .
\eea 
Alternatively, one could also consider $\phi^\dagger\cdot\Phi^2\Phi^1\cdot\phi$ as the secondary invariant.

Using the trace relations, we can argue that this set of invariants generates the complete space of gauge-invariant operators. For invariants that are traces of words constructed from $\Phi^1$ and $\Phi^2$ as letters, this argument was previously established in~\cite{deMelloKoch:2025ngs}. For invariants of the form $\phi^\dagger\cdot W\cdot\phi$, the argument is a straightforward extension of the discussion in Section~\ref{simpleexample}.

The trace relations also play a crucial role in demonstrating the necessity of introducing a secondary invariant and confirming that it is quadratically reducible. As discussed in Section~\ref{simpleexample}, the trace relations involving the vector fields can be constructed using the matrix \( \Psi \). Choosing \( A = \Phi^1 \), \( B = \Phi^2 \), and \( C = \Psi \), we obtain the following trace relation:
\bea
\phi^\dagger \cdot \Phi^1 \Phi^2 \cdot \phi + \phi^\dagger \cdot \Phi^2 \Phi^1 \cdot \phi &=& \Tr(\Phi^1) \phi^\dagger \cdot \Phi^2 \cdot \phi + \Tr(\Phi^2) \phi^\dagger \cdot \Phi^1 \cdot \phi + \phi^\dagger \cdot \phi \Tr(\Phi^1 \Phi^2)\cr\cr 
&&- \Tr(\Phi^1) \Tr(\Phi^2) \phi^\dagger \cdot \phi.
\eea
To determine the complete set of invariants involving two matrices and the vectors, we require a second relation to fix the difference \( \phi^\dagger \cdot \Phi^1 \Phi^2 \cdot \phi - \phi^\dagger \cdot \Phi^2 \Phi^1 \cdot \phi \). There is no such relation so we are forced to introduce either of these two terms as an invariant, explaining why the secondary invariant is necessary. Further analysis with the trace relations allows us to derive the identity
\bea
s^2&=&-{1\over 4} m_1^2 m_2^2 m_6^2+{1\over 4} m_2^2 m_3 m_6^2+{1\over 4} m_1^2 m_4 m_6^2-{1\over 4} m_3 m_4 m_6^2+{1\over 2} m_1 m_2^2 m_6 m_7-{1\over 2} m_1 m_4 m_6 m_7\cr\cr
&&-{1\over 2} m_2^2 m_7^2+{1\over 2} m_4 m_7^2+{1\over 2} m_1^2 m_2 m_6 m_8-{1\over 2} m_2 m_3 m_6 m_8 - m_5 m_7 m_8-{1\over 2} m_1^2 m_8^2+{1\over 2} m_3 m_8^2\cr\cr
&&- m_1 m_2 m_6 s+m_5 m_6 s+ m_2 m_7 s+ m_1 m_8 s
\eea
which is easy to verify using explicit numerical values for $\Phi^1$, $\Phi^2$ and $\phi$. This identity proves that $s^2$ can be replaced by a polynomial in the invariants with $s$ appearing at most linearly.

As a final example, consider a model with $d=2$, $f=2$, and $N=2$. The exact partition function is
\begin{align}
Z(x, y, z) &= \frac{1 + 4 x y z - 2 x^2 y^2 z + 4 x y z^2 - 4 x^3 y^3 z^2 + 2 x^2 y^2 z^3 - 4 x^3 y^3 z^3 - x^4 y^4 z^4}{(1 - x y)^4 (1 - z)^2 (1 - z^2)^3 (1 - x y z)^4},
\end{align}
which, again due to the presence of both positive and negative terms in the numerator, is not of Hironaka form. However, upon switching to the \emph{pure matrix grading} -- by identifying $x y \to z$ -- the partition function becomes
\begin{equation}
Z(z) = \frac{1 + 5 z^2 + 2 z^3 + 5 z^4 + z^6}{(1 - z)^6 (1 - z^2)^6},
\end{equation}
which \emph{is} of Hironaka form. This indicates that the ring of gauge-invariant operators is generated by 12 primary invariants, together with 13 non-trivial secondary invariants. The fact that the Hironaka structure is again recovered only after removing the vector/matrix grading highlights a rich interplay between matrix and vector degrees of freedom. The full structure of the invariant ring in mixed matrix/vector systems thus exhibits intricate features that merit further investigation.

To conclude this section, we present explicit counts of primary and secondary invariants for models with $d=2=f$, evaluated as a function of $N$. The results are summarized in Table~\ref{tab:growth} below. As in the previous examples, the partition function assumes the Hironaka form only when the pure matrix grading is employed. Notably, the number of secondary invariants grows extremely rapidly, in agreement with the behavior reported in~\cite{deMelloKoch:2025ngs}. This growth is sufficiently fast to account for the scaling expected from black hole entropy. Understanding the precise nature of this growth, especially its behavior for large values of $N$, remains an important open problem.

\begin{table}[h]
    \centering
    \begin{tabular}{|c|c|c|c|} 
        \hline
        $N$ & Primary Invariants & Secondary Invariants & Secondary Invariants for $f=0$\\ 
        \hline\hline
        2 & 12 & 14 &  1 \\ \hline
        3 & 21 & 2\,208  & 2 \\ \hline
        4 & 32 & 1\,684\,212 & 64 \\ \hline
        \hline
    \end{tabular}
    \caption{Growth in the number of primary and secondary invariants as $N$ increases for models with $d = 2$ and $f = 2$. For comparison, the final column lists the corresponding secondary invariant counts from~\cite{deMelloKoch:2025ngs} for a purely matrix model with $d = 2$ and $f = 0$. The identity operator is included in the count of secondary invariants.}\label{tab:growth}
\end{table}

\section{Discussion}\label{Discussion}

In this paper, we have studied the structure of the space of gauge-invariant operators at finite $N$, in coupled matrix–vector systems. For purely vector models with a $U(N)$ gauge symmetry and $f$ species of vectors, we find that when $f\leq N$, the space of gauge-invariant operators is freely generated by $f^2$ primary invariants. In contrast, when $f>N$, the generating set includes $N^2 + 2N(f - N)$ primary invariants along with a nontrivial set of secondary invariants. We have proposed closed-form analytic expressions for the number of secondary invariants, for arbitrary $f$ and $N$.

Our results have potential implications for higher-spin holography \cite{Klebanov:2002ja}. The free $U(N)$ vector model in $d$ dimensions is believed to be dual \cite{Giombi:2012ms} to a higher-spin gravity theory \cite{Vasiliev:1990en,Vasiliev:2003ev} on AdS$_{d+1}$ . Discretizing space onto a lattice with $N_l$ points, the construction of the full set of gauge-invariant operators maps directly onto the finite-dimensional problem we have studied, with $f=N_l$ vector species. Since any discretization of infinite space necessarily yields $f > N$, it follows that secondary invariants will be present in the generating set. Those secondary invariants with sufficiently high dimension are expected to correspond to new, nontrivial background geometries, while the primary invariants generate a Fock space of perturbative excitations propagating on these emergent backgrounds. From this point of view, the count of the number of secondary operators we have obtained look very natural. Indeed, we have $\log N_{\rm secondary}\approx f (2N\log 2)$. The factor of $f$, which is the number of lattice sites, measures the volume of the system so that we have obtained an extensive entropy. It would be interesting to make contact with thermal properties of vector models \cite{Shenker:2011zf} that depend on finite-$N$ effects. In the case of $O(N)$ vector models they become significant at temperatures $T \sim \sqrt{N}$. The basic gauge-invariant operators -- bilocals -- cease to be independent once their number exceeds $N$, necessitating a cutoff and leading to modified entropy scaling $S \sim N T^2 V$.

Our results suggest some interesting mathematical results that could be proved: one should prove that the ring of $U(N)$ invariants constructed from $f$ complex vectors, that are each $N$ dimensional, has Krull dimension equal to $N^2+2N(f-N)$. This should also be equal to the depth of the ring. It would also be interesting to compute the Krull dimension of the ring of invariants constructed from $f$ complex vectors and $d$ Hermitian matrices. For $f=1=d$ our results predict the Krull dimension is $2N$.

Although we have focused on free models, we expect our results to extend naturally to interacting theories. This is because trace relations are purely kinematical constraints on gauge-invariant quantities and remain valid, unchanged, in the presence of interactions. The trace relations are the sole input needed to settle key structural questions including whether a given set of invariants generates the full space of gauge-invariant operators, whether an invariant is quadratically reducible and what constraints exist among the invariants. Consequently, much of the structure of the gauge-invariant operator space at finite $N$ is largely insensitive to interactions.

Finally, as noted above, the invariants that generate the space of gauge-invariant operators may be interpreted as the collective fields in a collective field theory description. Establishing a precise connection between the results obtained in this work and the collective field theory formulation of matrix/vector models developed in~\cite{Avan:1995sp,Avan:1996vi,Aniceto:2006rr} is an interesting direction for future research.

\begin{center} 
{\bf Acknowledgements}
\end{center}
RdMK would like to thank Antal Jevicki for helpful discussions. This work is supported by a start up research fund of Huzhou University, a Zhejiang Province talent award and by a Changjiang Scholar award.

\begin{appendix}

\section{Bilinear Invariants and Gauge-Fixed Degrees of Freedom}\label{relatecount}

In this appendix, we consider a vector model with $f>N$ species of complex vectors. As discussed in Section~\ref{VMF}, gauge fixing allows us to express the $f$ complex vector fields in terms of $N^2 + 2N(f - N)$ independent real components. We now show that these independent components can be fully reconstructed from a set of $N^2 + 2N(f - N)$ bilinear gauge-invariant combinations. Thus, counting the independent components of the gauge-fixed vectors is equivalent to counting the number of independent bilinear invariants.

To make this argument concrete, we revisit the example from Equation~(\ref{gaugefixed}). We proceed step by step to show how each component of the gauge-fixed vectors can be recovered from suitable bilinear expressions.

We begin by noting that $r_1$ is uniquely determined by the invariant $\phi^{1\dagger} \cdot \phi^1 = r_1^2$, using the fact that $r_1>0$. Once $r_1$ is known, the bilinears $\phi^{1\dagger}\cdot\phi^2=r_1c_1$ and $\phi^{2\dagger}\cdot\phi^1=r_1 c_1^*$ allow us to determine both the real and imaginary parts of $c_1$. Applying the same logic, $c_2$ is determined by $\phi^{1\dagger} \cdot \phi^3$ and $\phi^{3\dagger}\cdot\phi^1$, $c_4$ by $\phi^{1\dagger}\cdot\phi^4$ and $\phi^{4\dagger}\cdot\phi^1$, $c_7$ by $\phi^{1\dagger}\cdot\phi^5$ and $\phi^{5\dagger}\cdot\phi^1$, $c_{11}$ by $\phi^{1\dagger}\cdot\phi^6$ and $\phi^{6\dagger}\cdot\phi^1$.

Next, since $c_1$ and $c_1^*$ are known, we can determine $r_2$ from the invariant
\bea
\phi^{2\dagger} \cdot \phi^2 = |c_1|^2 + r_2^2.
\eea
As above, $c_3$ is fixed by $\phi^{2\dagger}\cdot\phi^3$ and $\phi^{3\dagger}\cdot \phi^2$, $c_5$ by $\phi^{2\dagger}\cdot\phi^4$ and $\phi^{4\dagger}\cdot\phi^2$,
$c_8$ by $\phi^{2\dagger}\cdot\phi^5$ and $\phi^{5\dagger}\cdot\phi^2$,
$c_{12}$ by $\phi^{2\dagger}\cdot\phi^6$ and $\phi^{6\dagger}\cdot\phi^2$.

Proceeding similarly
\bea
\phi^{3\dagger} \cdot \phi^3 = |c_2|^2 + |c_3|^2 + r_3^2
\eea
determines $r_3$, and we find $c_6$ from $\phi^{3\dagger}\cdot\phi^4$ and $\phi^{4\dagger}\cdot\phi^3$, $c_9$ from $\phi^{3\dagger}\cdot\phi^5$ and $\phi^{5\dagger}\cdot\phi^3$, $c_{13}$ from $\phi^{3\dagger}\cdot\phi^6$ and $\phi^{6\dagger}\cdot\phi^3$.

Finally, since $c_4$, $c_5$, and $c_6$ are all known, we use:
\bea
\phi^{4\dagger} \cdot \phi^4 = |c_4|^2 + |c_5|^2 + |c_6|^2 + r_4^2
\eea
to determine \( r_4 \). The remaining components: $c_{10}$ and $c_{14}$ are obtained from $\phi^{4\dagger}\cdot\phi^5$, $\phi^{5\dagger}\cdot\phi^4$, $\phi^{4\dagger} \cdot \phi^6$, and $\phi^{6\dagger}\cdot\phi^4$.

Altogether, we have shown that the 32 real parameters remaining after gauge fixing can be reconstructed entirely from 32 independent bilinear gauge-invariant combinations. The generalization of this argument to arbitrary $N$ and $f$ follows immediately.

\end{appendix}

\end{document}